 \RequirePackage{fix-cm}

 \documentclass[smallextended]{svjour3} 

\def\theequation{\arabic{section}.\arabic{equation}}
\usepackage{amssymb,amsmath,amsfonts,latexsym} 
\usepackage{hyperref}
\usepackage{bm}

 \def\proofname{\smallskip \noindent\it Proof} 

\usepackage{mathtools}
\usepackage{multirow,array}
\usepackage{graphicx}
\usepackage{subfigure}
\usepackage{epstopdf}
\usepackage{float} 
\usepackage{color, xcolor}
\allowdisplaybreaks

\newcommand{\be}{\begin{equation}}
\newcommand{\ee}{\end{equation}}

\journalname{Fract. Calc. Appl. Anal.} 

\begin{document}

  
\title{Asymptotic cycles in fractional generalizations of multidimensional maps}

\titlerunning{Asymptotic cycles in fractional generalizations \dots}

\author{
        Mark Edelman$^1$ 
 }

\authorrunning{M. Edelman} 
\institute{Mark Edelman$^{1,*}$
\at
Department of Physics, Stern College at Yeshiva University \\
245 Lexington Ave, New York, NY 10016, USA \\
Courant Institute of Mathematical Sciences, New York University \\
251 Mercer St., New York, NY 10012, USA \\
\email{edelman@cims.nyu.edu} $^*$ corresponding author 
}

\date{Received: ......  / Revised: ...... / Accepted: ......}


\maketitle

\begin{abstract}

In regular dynamics, discrete maps are model presentations of discrete dynamical systems, and they may approximate continuous dynamical systems. Maps are used to investigate general properties of dynamical systems and to model various natural and socioeconomic systems. They are also used in engineering. Many natural and almost all socioeconomic systems possess memory which, in many cases, is power-law-like memory. Generalized fractional maps, in which memory is not exactly the power-law memory but the asymptotically power-law-like memory, are used to model and investigate general properties of these systems. 

In this paper we extend the definition of the notion of generalized fractional maps of arbitrary positive orders that previously was defined only for maps which, in the case of integer orders, converge to area/volume-preserving maps. Fractional generalizations of H\'enon and Lozi maps belong to the newly defined class of generalized fractional maps. We derive the equations which define periodic points in generalized fractional maps. We consider applications of our results to the fractional and fractional difference H\'enon and Lozi maps.



\end{abstract} 


\section{Introduction} \label{sec:1}

\setcounter{section}{1} \setcounter{equation}{0} 

Many natural and socioeconomic systems possess memory. Systems with memory are also widely used in engineering. In many cases, this memory is a power-law-like memory. Maps with memory may be used to describe discrete systems with memory, to model continuous systems with memory, and all numerical schemes to solve fractional differential equations may be considered as maps with memory. A review of the economic applications of maps with memory may be found in \cite{T1} (see also \cite{T2,T3}).
Fractional maps allow modeling of the distributions of lifespans of living species consistent with the Gompertz distribution and a limited lifespan \cite{ME11}. There is an abundance of papers in which fractional maps are used to model memristors (see, e.g, \cite{memr}), to encrypt images and signals (see, e.g. \cite{crypto1,crypto2}), to model spread of infections (see, e.g., \cite{epi}), to control systems (see, e.g., \cite{Ost,Ort}), etc.

The first maps with memory introduced and investigated in scientific publications were not fractional - the memory was directly introduced in the forms which were relevant to the physical formulation of the corresponding problems (see \cite{MM1,MM2,MM3,MM4,MM5,MM6}). In the one-dimensional case, maps with memory may be written as
\begin{equation}
x_{n+1}=\sum^{n}_{k=0}V_{\alpha}(n,k)G_K(x_k),
\label{LTM}
\end{equation}
where $V_{\alpha}(n,k)$ and $\alpha$ characterize memory effects
and $K$ is a parameter.
In many cases, maps are convolutions with $V_{\alpha}(n,k)=V_{\alpha}(n-k)$. The initial applications were related to the case of exponentially decaying memory. It has been shown in \cite{ME1} that directly introduced maps with power-law memory ($V_{\alpha}(n,k)=(n-k)^{\alpha-1}$ coincide with the fractional maps introduced in \cite{ZT1} as solutions of fractional equations describing kicked systems. In the case of the fractional maps with time step $h$, corresponding fractional differential equations are recovered in the limit $h \to \infty$ (see \cite{ME1}). In 2006, Stanislavsky investigated the logistic map with memory, whose kernel was taken from an algorithm of the numerical fractional integration \cite{MM7}. Fractional maps, introduced in \cite{ZT1}, were initially investigated in \cite{ME2,ME3,ME4,ME5,ME6,XML}.

Discrete fractional calculus was proposed in \cite{GZ,MR} (see also more recent reviews \cite{Ost,Ort,Agr,GP,HBV2,HBV4}). Solutions of fractional difference equations were initially investigated in 
\cite{AE1,Anast1,AE2,Anast2,BFT}. In \cite{CLZ,WB1,WB2,WB2C}, the authors showed that solutions of fractional difference equations may be written as fractional (we will call them fractional difference) maps with the falling factorial memory.

It soon became obvious that fractional and fractional difference maps may be considered as particular forms of a wider class of maps which possess similar properties \cite{ME11,HBV2,HBV4,ME7,ME8,ME9,ME10}. The notion of the generalized fractional maps (GFM) was introduced in \cite{ME12}, in which the equations defining asymptotically periodic points were derived. The definition of GFM was further considered in \cite{ME13,ME14,ME15}, in which the authors investigated maps' asymptotically periodic points, stability, and bifurcations. 

GFM are particular forms of the Volterra difference equations of convolution type (see, e.g. Chapter 6 in \cite{Elaydi}) with kernels, whose differences are absolutely summable, but the series of kernels are diverging. Proved in \cite{ME16} Theorem 1.1 is the most important property of the GFM's kernels, which allows to investigate the asymptotic behavior of GFM. 

GFM introduced in the above cited papers are maps which, in the case of integer orders, produce area/volume preserving maps. In this paper we extend the notion of GFM to include multi-dimensional maps, like H\'enon and Lozi maps, which are not area/volume preserving.

\section{Preliminaries} \label{sec:2}

\setcounter{section}{2} \setcounter{equation}{0}

As we mentioned in Section \ref{sec:1}, Theorem 1.1 of \cite{ME16} is the most important property of the GFM's kernels. It is formulated as the following:

\begin{theorem}\label{Th1} 
If 
\begin{eqnarray}
\lim_{k \rightarrow \infty}x_k=X,   
\  \ 	 
\lim_{N \rightarrow \infty}\sum_{k=0}^N a_k=S,
\label{The1_1}
\end{eqnarray}
and the series is converging absolutely ($\{a_k\} \in l^1$), then 
\begin{eqnarray}
\Sigma= \lim_{N \rightarrow \infty}\sum_{k=0}^N x_{N-k}a_k=XS,
\label{The1_2}
\end{eqnarray} 
where $k,N \in \mathbb{N}_0$ and $x_k,a_k,X,S,\Sigma \in \mathbb{R}$. 

If $S=\infty$, and $\Sigma$ is finite, then $X=0$.
\end{theorem}

\subsection{\bf Regular H\'enon, Lozy, and $p$-dimensional maps }
\label{sec:2.1}

The H\'enon map \cite{Henon}, written as a difference equation, is:
{ \setlength\arraycolsep{0.5pt}
\begin{eqnarray}
&&x_{n+1}-x_n=1-x_n-ax_n^2+y_n
, \ \ \ \  \\
&&y_{n+1}-y_n=bx_n-y_n.
\label{Henon}
\end{eqnarray}
}
In the classical H\'enon map, the parameter values are: $a = 1.4$ and $b = 0.3$.

The Lozi map \cite{Lozi}, written as a difference equation, is:
{ \setlength\arraycolsep{0.5pt}
\begin{eqnarray}
&&x_{n+1}-x_n=1-x_n-a|x_n|+y_n
, \ \ \ \  \\
&&y_{n+1}-y_n=bx_n-y_n.
\label{Lozi}
\end{eqnarray}
}

In the general case of $p$-dimensional maps, the equations can be written as 
{ \setlength\arraycolsep{0.5pt}
\begin{eqnarray}
&&\Delta x_{i}(t)=-G_{i,K}(x_{1}(t),x_{2}(t),...,x_{p}(t)), \  \ 1 \le i \le p, 
\  \ t\in \mathbb{N},
\label{nDim_t}
\end{eqnarray}
}
where the forward difference operator $\Delta$ is defined as
\begin{equation}
\Delta x(n)= x(n+1)-x(n),
\label{Delta}
\end{equation}
with the initial conditions $x_i(0)=x_{i,0}$. These equations, with $x_{i,n}=x_{i}(n)$ for $t=n$, can be written as
{ \setlength\arraycolsep{0.5pt}
\begin{eqnarray}
&&\Delta x_{i,n}=-G_{i,K}(x_{1,n},x_{2,n},...,x_{p,n}), \  \ 1 \le i \le p, 
\  \ n \ge 0,
\label{nDim}
\end{eqnarray}
}
which are equivalent to the map:
{ \setlength\arraycolsep{0.5pt}
\begin{eqnarray}
&&x_{i,n+1}=F_{i,K} (x_{1,n},x_{2,n},...,x_{p,n}), \  \ 1 \le i \le p, \  \ n \ge 0 
\label{nDimBase}
\end{eqnarray}
}
with 
{ \setlength\arraycolsep{0.5pt}
\begin{eqnarray}
&&G_{i,K}(x_{1,n},x_{2,n},...,x_{p,n})= x_{i,n}- F_{i,K}(x_{1,n},x_{2,n},...,x_{p,n}), 
\label{Gdef}
\end{eqnarray}
}
where $K$ stands for a set of parameters.

\subsection{\bf Definition of fractional and fractional difference maps}
\label{sec:2.2}

The system Eq.~(\ref{nDim_t}) turns into a system of fractional         Caputo delta $h$-difference equations when the first differences $\Delta x_{i}(t)$ are substituted by the orders $0<\alpha_i<1$ Caputo forward $h$-differences (see review \cite{HBV2} and references therein)
\begin{eqnarray}
&&(_0\Delta^{\alpha_i}_{h,*} x_i)(t) = 
\nonumber \\
&&-G_{i,K}(x_1(t+(\alpha_i-1)h), x_2(t+(\alpha_i-1)h, ..., x_p(t+(\alpha_i-1)h),
\label{LemmaDif_n_h}
\end{eqnarray}
where $t\in \{h\mathbb{N}_{1}\}$, with the initial conditions 
 \begin{equation}
x_i(0) = x_{i,0}, \ \ 1 \le i \le p, \   \ 
\label{LemmaDifICn_h}
\end{equation}
which is equivalent to the system of Volterra difference equations of convolution type (see Theorem 9 in \cite{HBV2}): 
\begin{eqnarray} 
&&x_{i,n+1} = x_{i,0} 
\nonumber \\
&&-\frac{h^{\alpha_i}}{\Gamma(\alpha_i)}  
\sum^{n}_{s=0}(n-s-1+\alpha_i)^{(\alpha_i-1)} 
G_{i,K}(x_{1,s},x_{2,s},..., x_{p,s}), 
\label{FalFacMap_hN}
\end{eqnarray}
where $x_{i,k}=x_i(kh)$. 
The definition of the falling factorial $t^{(\alpha)}$ is
\begin{equation}
t^{(\alpha)} =\frac{\Gamma(t+1)}{\Gamma(t+1-\alpha)}, \ \ t\ne -1, -2, -3.
...
\label{FrFacN}
\end{equation}
The falling factorial is asymptotically a power function:
\begin{equation}
\lim_{t \rightarrow
  \infty}\frac{\Gamma(t+1)}{\Gamma(t+1-\alpha)t^{\alpha}}=1,  
\ \ \ \alpha \in  \mathbb{R}.
\label{GammaLimitN}
\end{equation}
The $h$-falling factorial $t^{(\alpha)}_h$ is defined as
\begin{eqnarray}
&&t^{(\alpha)}_h =h^{\alpha}\frac{\Gamma(\frac{t}{h}+1)}{\Gamma(\frac{t}{h}+1-\alpha)}= h^{\alpha}\Bigl(\frac{t}{h}\Bigr)^{(\alpha)},
\nonumber \\
&&\frac{t}{h} \ne -1, -2, -3,
....
\label{hFrFacN}
\end{eqnarray}
After substitution 
\begin{equation}
{G}_i^0(x_{1,s},x_{2,s},..., x_{p,s})=h^\alpha_i G_{i,K}(x_{1,s},x_{2,s},..., x_{p,s})/\Gamma(\alpha_i), 
\label{Sub1}
\end{equation}
assuming 
{\setlength\arraycolsep{0.5pt}
\begin{eqnarray}
&&U_{\alpha}(n)=(n+\alpha-2)^{(\alpha-1)} 
, \ \ \ \  \nonumber \\  
&&U_{\alpha}(1)=(\alpha-1)^{(\alpha-1)}=\Gamma(\alpha),
\label{UnFrDifN}
\end{eqnarray} 
the system Eq.~(\ref{FalFacMap_hN}) may be written as 
\begin{eqnarray}
x_{i,n} = x_{i,0}  
-\sum^{n-1}_{k=0} {G}_i^0(x_{1,k},x_{2,k},..., x_{p,k}) U_{\alpha_i}(n-k).
\label{FrUUMapN}
\end{eqnarray} 
With the kernel Eq.~(\ref{UnFrDifN}), this system turns into a fractional difference $p$-dimensional map. When the kernel is (see \cite{ME14}) 
\begin{eqnarray}
U_{\alpha}(n)=n^{\alpha-1}, \ \ \ \  U_{\alpha}(1)=1,
\label{UnFrN}
\end{eqnarray} 
the system may be called a fractional $p$-dimensional map. 

We define the map as a generalized fractional $p$-dimensional map when the kernel $U_{\alpha}(n)$ belongs to the space of functions 
$\mathbb{D}^0(\mathbb{N}_1)$ defined as (see \cite{ME13})
{\setlength\arraycolsep{0.5pt}
\begin{eqnarray}
&&\mathbb{D}^i(\mathbb{N}_1) \ \ = \ \ \{f: \left|\sum^{\infty}_{k=1}\Delta^if(k)\right|>N, 
\nonumber \\
&& \forall N, \ \ N \in \mathbb{N}, 
\sum^{\infty}_{k=1}\left|\Delta^{i+1}f(k)\right|=C, \ \ C \in \mathbb{R}_+\}.
\label{DefForm}
\end{eqnarray}
}
We also assume that $U_{\alpha}(n)=0$ for $n<1$.
For any real number $a$, $\mathbb{N}_a=\{a, a+1, a+2, a+3, \ldots\}$, and $\mathbb{N}=\mathbb{N}_0$.

\section{Periodic Points} \label{sec:3} 
\setcounter{section}{3} \setcounter{equation}{0}

Now we may transform Eq.~(\ref{FrUUMapN}), following the same steps as in Eq.~(10) from \cite{ME14}. Easily verifiable result for any $0<m<l$ is 
{\setlength\arraycolsep{0.5pt}   
\begin{eqnarray} 
&&x_{i,lN+m+1}-x_{i,lN+m}
\nonumber \\
&&=\sum^{N-1}_{k=0}\sum^{l-1}_{j=1} G_i^0(x_{1,lN+m-lk-j},x_{2,lN+m-lk-j}, ..., x_{p,lN+m-lk-j})\Bigl[
U_{\alpha_i} (lk+j)
\nonumber \\
&& - U_{\alpha_i} (lk+j+1)\Bigr]+\sum^{N-1}_{k=1} G_i^0(x_{1,lN+m-lk}, x_{2,lN+m-lk}, ..., x_{p,lN+m-lk})\Bigl[
U_{\alpha_i} (lk)
\nonumber \\
&& - U_{\alpha_i} (lk+1)\Bigr]- G_i^0(x_{1,lN+m},x_{2,lN+m}, ..., x_{p,lN+m})U_{\alpha_i} (1)
\nonumber \\
&&+ G_i^0(x_{1,m}, x_{2,m}, ..., x_{p,m})U_{\alpha} (lN) +S_{i,0},
\label{mth_lCycle_point_dif}
\end{eqnarray}
}
where $S_{i,0}$ is a sum of a finite number of elements, which tends to zero as $N \rightarrow \infty$.  Let's assume that in the limit $N \rightarrow \infty$ the system converges to a period $l$ sink ($l$-cycle)
 \begin{equation}
x_{i,l,m}=\lim_{N \rightarrow
  \infty} x_{i,Nl+m}, \  \  \ 0<m<l+1,
\label{TlpointEx}
\end{equation}
and consider the limit of Eq.~(\ref{mth_lCycle_point_dif}) as  
$N \rightarrow \infty$.
Following the same steps as in Eq.~(12) -- Eq.~(18) from \cite{ME14} and using the Theorem \ref{Th1}, we come to the following system of $(l-1)\times p$ equations:
{\setlength\arraycolsep{0.5pt}   
\begin{eqnarray} 
&&x_{i,l,m+1}-x_{i,l,m}= \sum^{m-1}_{j=0}S_{i,j+1,l} G_i^0(x_{1,l,m-j},x_{2,l,m-j}, ..., x_{p,l,m-j})
\nonumber \\
&&+\sum^{l-1}_{j=m}S_{i,j+1,l} G_i^0(x_{1,l,m-j+l},x_{2,l,m-j+l}, ..., x_{p,l,m-j+l}), 
\nonumber \\
&&0<m<l, \  \ 0<i \le p,
\label{LimDifferencesN}
\end{eqnarray}
}
where 
\begin{equation}
{S}_{i,j+1,l}=\sum^{\infty}_{k=0}\Bigl[
U_{\alpha_i} (lk+j) - U_{\alpha_i} (lk+j+1)\Bigr] , \  \ 0 \le j<l.
\label{S1}
\end{equation}
It is easy to see that 
\begin{equation}
\sum^{l}_{j=1}S_{i,j,l}=0 , \  \ 0 < i \le p.
\label{Ssum}
\end{equation}
As in \cite{ME14} (Eq.~(20) -- Eq.~(23)), the limits when $N \rightarrow \infty$ of the totals of the projections of the periodic points 
{\setlength\arraycolsep{0.5pt}   
\begin{eqnarray} 
&& \sum^{l}_{m=1}x_{i,lN+m}=lx_{i,0}
\nonumber \\
&& -\sum^{l}_{m=1}\sum^{l}_{j=1}\sum^{N-1}_{k=0} G_i^0(x_{1,lN+m-lk-j}, x_{2,lN+m-lk-j}, ..., x_{p,lN+m-lk-j})U_{\alpha_i} (lk+j)
\nonumber \\
&&-\sum^{l}_{m=1}\sum^{m}_{j=1} G_i^0(x_{1,m-j}, x_{2,m-j}, ..., x_{p,m-j}) U_{\alpha
_i} (lN+j),
\label{lCycle_pointTotal}
\end{eqnarray}
}
after application of the second statement of Theorem \ref{Th1}, give additional $p$ equations:
\begin{equation}
\sum^{l}_{j=1} G_i^0(x_{1,l,j},x_{2,l,j}, ..., x_{p,l,j})=0, \ 0 < i \le p,
\label{Close}
\end{equation}
which together with Eq.~(\ref{LimDifferencesN}) define $l$ ($0< m \le l$) $p$-dimensional periodic points of the $l$-cycle $x_{i,l,m}$.

\section{Generalized fractional H\'enon map} \label{sec:4} 
\setcounter{section}{4} \setcounter{equation}{0}

In the Generalized fractional H\'enon map, the functions generating the map are:
{ \setlength\arraycolsep{0.5pt}
\begin{eqnarray}
&&G_{1H}^0=-\frac{h^{\alpha_1}}{\Gamma(\alpha_1)}(1-x_{1,n}-ax_{1,n}^2+ x_{2,n})
, \ \ \ \  
\nonumber \\
&&G_{2H}^0= -\frac{h^{\alpha_2}}{\Gamma(\alpha_2)} (bx_{1,n}-x_{2,n}).
\label{GFrHenon}
\end{eqnarray}
}
The fixed points of the generalized fractional H\'enon map 
are the same as the fixed points of the classical H\'enon map:
{ \setlength\arraycolsep{0.5pt}
\begin{eqnarray}
&&x_{1f}=\frac{b-1 \pm \sqrt{(b-1)^2+4a}}{2a},
\nonumber \\
&&x_{2f}=b x_{1f}.
\label{FixedH}
\end{eqnarray}
}

\subsection{\bf The fractional H\'enon map} \label{sec:4_1} 

Dissipative 2D maps may be introduced as solutions of differential equations with dissipation and kicks. Corresponding fractional maps may be obtained as solutions of the dissipative differential equations with fractional derivatives (see \cite{FDSM,TarasovHenon}). The generalized fractional dissipative map is obtained in \cite{TarasovHenon} as a solution of the equation
\begin{equation}
_0D^{\alpha}_tx(t)-q _0D^{\alpha-1}_tx(t) =KG(x(t)) \sum^{\infty}_{n=0} \delta(t-nT),  
 \    \ \ 1<\alpha \le 2.
\label{UDFMRL}
\end{equation}
Then, the fractional H\'enon map is obtained by assuming 
\begin{equation}
G(x)=-\frac{q}{1+b} [1 + (1 + b)x + ax^2]
\label{TarHenon}
\end{equation}
and integrating Eq.~(\ref{UDFMRL}).

We should note that the technique to derive fractional maps from differential equations of the orders $\alpha \le 1$ with kicks was developed in 2013 (see \cite{ME5,ME6}). So, the way in which Tarasov \cite{TarasovHenon} generalized the H\'enon map was the most appropriate in 2010. The fractional H\'enon map, which is a particular form of the generalized fractional map proposed in this paper, is a solution of the system of two Caputo fractional differential equations of the orders $0 < \alpha_i \le 1$, $i=1,2$ with kicks (see Section 2.3 in \cite{ME6} and Section 3 in \cite{HBV2}):   
\begin{equation}
_0^CD^{\alpha_i}_tx_i(t) +G_{i,K}(x_1(t- \Delta ), x_2(t- \Delta )) \sum^{\infty}_{n=-\infty} \delta \Bigl(\frac{t}{h}-(n+\varepsilon)
\Bigr)=0,    
\label{UM1D2DdifC}
\end{equation}
 where $\varepsilon > \Delta > 0$, $\varepsilon  \rightarrow 0$, $0 < \alpha_i \le 1$,  $\alpha_i \in \mathbb{R}$, index $K$ stands for the set of parameters, and the initial conditions 
\begin{equation}
x_i(0+)=b_{i,0},  \    \ i=1,2.
\label{UM1D2DdifCic}
\end{equation}
The left-sided Caputo fractional derivative $_0^CD^{\alpha}_t x(t)$ is defined for $t>0$ \cite{KST,Podlubny,SKM} as
\begin{equation}
_0^CD^{\alpha}_t x(t)=_0I^{n-\alpha}_t \ D^n_t x(t) =
\frac{1}{\Gamma(n-\alpha)}  \int^{t}_0 
\frac{ D^n_{\tau}x(\tau) d \tau}{(t-\tau)^{\alpha-n+1}},
\label{Cap}
\end{equation}
where $n-1 <\alpha \le n$.
The system of equations Eq.~(\ref{UM1D2DdifC}) is equivalent to the 2-D map 
{\setlength\arraycolsep{0.5pt}
\begin{equation}
x_{i,n+1}= x_{i,0} 
-\frac{h^{\alpha_i}}{\Gamma(\alpha_i)}\sum^{n}_{k=0} G_{i,K}(x_{1,k}, x_{2,K}) (n-k+1)^{\alpha_i-1},
\label{FrCMapx}
\end{equation}
}
The fractional H\'enon map is obtained when
{ \setlength\arraycolsep{0.5pt}
\begin{eqnarray}
&&G_{1,K}=-(1-x_{1,n}-ax_{1,n}^2+ x_{2,n})
, \ \ \ \  
\nonumber \\
&&G_{2,K}= -(bx_{1,n}-x_{2,n}),
\label{FracHenon}
\end{eqnarray}
}
which corresponds to the definition of the generalized fractional H\'enon map Eq.~(\ref{GFrHenon}) with
{\setlength\arraycolsep{0.5pt}
\begin{equation}
G_{iH}^0=-\frac{h^{\alpha_1}}{\Gamma(\alpha_1)} G_{i,K}, \   \ i=1,2.
\label{Gs}
\end{equation}
}

The numerical investigation of the fractional H\'enon map is beyond the scope of this paper and will be performed elsewhere.

\subsection{\bf Period two points of the fractional difference H\'enon map } \label{sec:4_2} 

In this section, we will assume $\alpha_1=\alpha_2=\alpha$. Then (see Eq.~(31) from \cite{ME15}),
{ \setlength\arraycolsep{0.5pt}
\begin{eqnarray}
&&S_{1,2,2}=S_{2,2,2}=\Gamma(\alpha)2^{-\alpha},
\nonumber \\
&&S_{1,1,2}=S_{2,1,2}=-\Gamma(\alpha)2^{-\alpha},
\label{S2n}
\end{eqnarray}
}
and the equations defining $T=2$ points $\{(x_{1,2,1}, x_{2,2,1}),(x_{1,2,2}, x_{2,2,2})\}$ may be written as 
{ \setlength\arraycolsep{0.5pt}
\begin{eqnarray}
&&x_{1,2,2}-x_{1,2,1}=(h/2)^{\alpha}
\Bigl\{ (x_{1,2,2}-x_{1,2,1})[1+a(x_{1,2,2}+x_{1,2,1})]
\nonumber \\
&& -(x_{2,2,2}-x_{2,2,1})\Bigr\},
\nonumber \\
&& x_{2,2,2}-x_{2,2,1}=(h/2)^{\alpha}
[(x_{2,2,2}-x_{2,2,1})-b(x_{1,2,2}-x_{1,2,1})],
\nonumber \\
&&2- (x_{1,2,2}+x_{1,2,1})-a(x_{1,2,2}^2+x_{1,2,1}^2)+ (x_{2,2,2}+x_{2,2,1})=0,
\nonumber \\
&&b(x_{1,2,1}+x_{1,2,2})-(x_{2,2,1}+x_{2,2,2})=0.
\label{T2n}
\end{eqnarray}
}

If we exclude the fixed points, then the solution of these equations defining two $T=2$ points can be written as
{ \setlength\arraycolsep{0.5pt}
\begin{eqnarray}
&&x_{1,2,1} = \frac{C_1 \pm \sqrt{2C_2-C_1^2}}{2},
\nonumber \\
&&x_{2,2,1}=\frac{b}{2}C_1+\frac{b}{C_0}(x_{1,2,1}-\frac{C_1}{2}),
\nonumber \\
&&x_{1,2,2}=C_1-x_{1,2,1},
\nonumber \\
&& x_{2,2,2}=bC_1-x_{2,2,1},
\label{T2nH}
\end{eqnarray}
}
where 
{ \setlength\arraycolsep{0.5pt}
\begin{eqnarray}
&&C_0=1-(2/h)^{\alpha},
\nonumber \\
&&C_1=\frac{b-C_0^2}{aC_0},
\nonumber \\
&&C_2=\frac{1}{a}[2+(b-1)C_1].
\label{Cs}
\end{eqnarray}
}
When $h=1$ and $\alpha=1$, we recover two $T=2$ points of the classical H\'enon map:
{ \setlength\arraycolsep{0.5pt}
\begin{eqnarray}
&&x_{1,2} = \frac{1-b \pm \sqrt{4a-3(1-b)^2}}{2a},
\nonumber \\
&& y_{1,2}=bx_{2,1}.
\label{T2HC}
\end{eqnarray}
}
The fixed point - $T=2$ point bifurcation (we assume $x$ of $a$ bifurcation diagram with the fixed $b$) occurs when 
{ \setlength\arraycolsep{0.5pt}
\begin{eqnarray}
&&x_{1,1bif} = \frac{b-C_0^2}{2aC_0},
\nonumber \\
&&x_{2,1bif}=bx_{1,1bif},
\nonumber \\
&&a_{bif}=\frac{b-C_0^2}{4C_0}\Bigl[\frac{b}{C_0}-C_0-2(b-1)\Bigr].
\label{Bif1H}
\end{eqnarray}
}

To the best of our knowledge, the only paper in which bifurcation diagrams of the fractional H\'enon map were drawn is the article \cite{Hu}. The diagrams ($x_1$ as a function of $a$) were obtained for four different values of parameters $b$ and $\alpha$ including the values $b=0.05$ and $\alpha=0.6$ for which the fixed and $T=2$ points (stable and unstable) are depicted in Fig.~\ref{fig1}. The value of $a$ at the bifurcation point $a_{bif}=0.243$ in Fig.~\ref{fig1} is the same value as the value calculated from Eq.~(\ref{Bif1H}). The two fixed points in Fig.~\ref{fig1} are drawn only for values $a<a_{bif}$ because, according to the numerical calculations form \cite{Hu}, they are unstable for those $a$. More than that, it is well known that the solution of Eq.~(\ref{FixedH}) with the negative sign, which does not bifurcate, is always unstable. From the comparison of the figure from \cite{Hu} and Fig.~\ref{fig1}, it is obvious that the accuracy of calculations using formulae obtained in this paper is much better than from the direct map simulations.
\begin{figure}[!t]
\includegraphics[width=1.0 \textwidth]{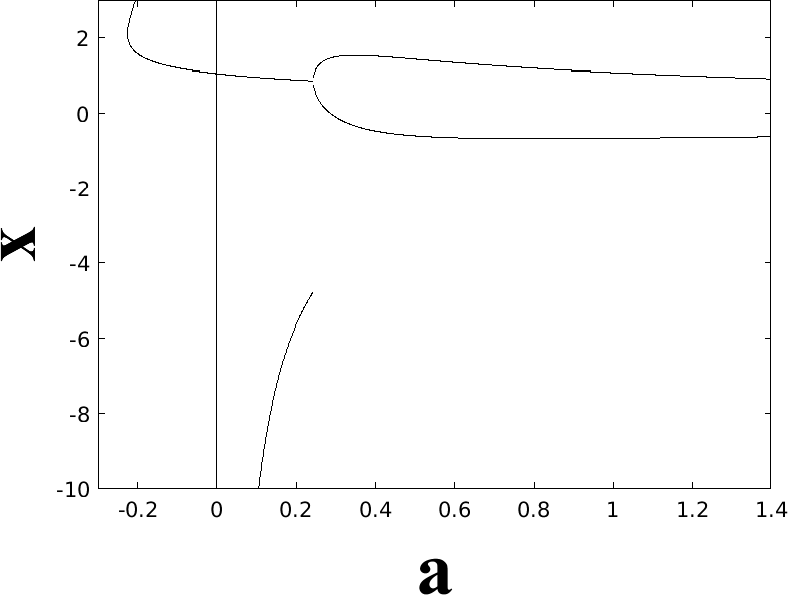}
\vspace{-0.25cm}
\caption{Fixed and $T=2$ points (stable and unstable) for the 
fractional difference H\'enon map with $b=0.05$ and $\alpha=0.6$.}
\label{fig1}
\end{figure}


\section{Fractional difference Lozi map} \label{sec:5} 
\setcounter{section}{5} \setcounter{equation}{0}

In the fractional difference Lozi map, the functions generating the map are:
{ \setlength\arraycolsep{0.5pt}
\begin{eqnarray}
&&G_{1H}^0=-\frac{h^{\alpha_1}}{\Gamma(\alpha_1)}(1-x_{1,n}-a|x_{1,n}|+ x_{2,n})
, \ \ \ \  \\
&&G_{2H}^0= -\frac{h^{\alpha_2}}{\Gamma(\alpha_2)} (bx_{1,n}-x_{2,n}).
\label{GFrLozi}
\end{eqnarray}
}
The fixed points of the generalized fractional Lozi map 
are the same as the fixed points of the classical Lozi map:
{ \setlength\arraycolsep{0.5pt}
\begin{eqnarray}
&&x_{1f}=\frac{1}{1+a-b},
\nonumber \\
&&x_{2f}=b x_{1f}
\label{FixedL1}
\end{eqnarray}
}
if $x_{1f}>0$ and
{ \setlength\arraycolsep{0.5pt}
\begin{eqnarray}
&&x_{1f}=\frac{1}{1-a-b},
\nonumber \\
&&x_{2f}=b x_{1f}
\label{FixedL2}
\end{eqnarray}
}
if $x_{1f} \le 0$.
The equations defining period two points, in the case $\alpha_1=\alpha_2=\alpha$ may be obtained the same way as they were obtained in Section~\ref{sec:4_2} for the H\'enon map: 
\begin{eqnarray}
&&x_{1,2,2}-x_{1,2,1}=(h/2)^{\alpha}
\Bigl\{x_{1,2,2}-x_{1,2,1}+a(|x_{1,2,2}|-|x_{1,2,1}|)
\nonumber \\
&& -(x_{2,2,2}-x_{2,2,1})\Bigr\},
\nonumber \\
&& x_{2,2,2}-x_{2,2,1}=(h/2)^{\alpha}
[(x_{2,2,2}-x_{2,2,1})-b(x_{1,2,2}-x_{1,2,1})],
\nonumber \\
&&2- (x_{1,2,2}+x_{1,2,1})-a(|x_{1,2,2}|+|x_{1,2,1}|)+ (x_{2,2,2}+x_{2,2,1})=0,
\nonumber \\
&&b(x_{1,2,1}+x_{1,2,2})-(x_{2,2,1}+x_{2,2,2})=0.
\label{T2nLozi}
\end{eqnarray}
}

Stability and bifurcations in the regular Lozi map were investigated in \cite{LoziBif}, and some numerical results for the fractional Lozi map were obtained in \cite{FrLoziBif}. We will not calculate the Lozi's map periodic points and analyze their stability here -- it will be done in one of the following publications.

\section{Conclusion} \label{sec:6} 

In this paper we defined fractional generalization of multidimensional maps and derived equations defining their periodic points. This allows calculations of periodic points in a large set of maps, which are already used in publications, including the H\'enon and Lozi maps. The problems of stability and bifurcations in these maps will be investigated in publications that will follow.


\begin{acknowledgements}
The author acknowledges continuing support from Yeshiva University 
and expresses his gratitude to the administration of Courant
Institute of Mathematical Sciences at NYU
for the opportunity to perform computations at Courant.  \end{acknowledgements}

 \section*{\small
 Conflict of interest} 

 {\small
 The author declares that he has
 no conflict of interest.}




\bigskip  

\small 
\noindent
{\bf Publisher's Note}
Springer Nature remains neutral with regard to jurisdictional claims in published maps and institutional affiliations.


\begin{thebibliography} {99}
 \normalsize 



\bibitem {T1} Tarasov, V.E., Tarasova, V.V.: Economic Dynamics with Memory: Fractional Calculus Approach. De Gruyter, Berlin, Boston (2021). DOI: 10.1515/9783110627459

\bibitem{T2} Tarasov, V.E., Tarasova, V.V.: Long and short memory in economics: fractional-order difference and differentiation.
Int. J. Management Social Sciences \textbf{5}, 327--334 (2016).
DOI: 10.21013/jmss.v5.n2.p10

\bibitem{T3} Tarasova, V.V., Tarasov, V.E.: Logistic map with memory from economic model. Chaos, Solitons and Fractals \textbf{95}, 84--91 (2017)

\bibitem {ME11} Edelman, M.:
Evolution of systems with power-law memory: Do we have to die?
(Dedicated to the Memory of Valentin Afraimovich). In: Skiadas, C.H., Skiadas C. (eds.)
{Demography of Population Health, Aging and Health Expenditures}, 
pp. 65--85. Springer, eBook (2020)

\bibitem {memr} He, S., Sun, K., Peng, Y., Wang, L.:
Wang Modeling of discrete fracmemristor and its application.
AIP Advances \textbf{10}, 015332 (2020)


\bibitem {crypto1} Huang, L.-L., Baleanu, D, Wu, G.-C., Zeng, S.-D.:
A new application of the fractional logistic map.
Romanian Journal of Physics \textbf{61}, 1172--1179 (2016)

\bibitem {crypto2} Ding, D., Wang, J., Wang, M. et al. Controllable multistability of fractional-order memristive coupled chaotic map and its application in medical image encryption. Eur. Phys. J. Plus \textbf{138}, 908 (2023).

\bibitem {epi} Selvam1, A.G.M., Vianny, D.A.: Discrete Fractional Order SIR Epidemic Model of Childhood Diseases with Constant Vaccination and it's Stability. Int. J. Tech. Innova. Mod. Engin. Sci. \textbf{4}, 405--410 (2018)

\bibitem {Ost} Ostalczyk P.: Discrete Fractional Calculus: Applications in Control and Image Processing. World Scientific, River Edge, NJ (2016)

\bibitem {Ort} Ortigueira, M.: Discrete-time fractional difference calculus: origins, evolutions, and new formalisms.
Fractal Fract. \textbf{7}, 502 (2023)


\bibitem {MM1} Fulinski, A., Kleczkowski, A.S.:
Nonlinear maps with memory.
{Physica Scripta} \textbf{335}, 119--122 (1987)

\bibitem {MM2} Fick, E, Fick, M., Hausmann, G.:
Logistic equation with memory,.
{Physical Review A} \textbf{44}, 2469--2473 (1991)
Simulating memory effects with discrete

\bibitem {MM3} Giona, M.:
Dynamics and relaxation properties of complex systems with memory.
{1Nonlinearity} \textbf{14}, 911--925 (1991)

\bibitem {MM4} Hartwich, K., Fick, E.:
Hopf bifurcations in the logistic map with oscillating memory. 
{Physics Letters A} \textbf{177}, 305--310 (1993)

\bibitem {MM5} Gallas, J.A.C.:
 dynamical systems. 
{Physica A} \textbf{195}, 417--430 (1993)

\bibitem {MM6} Gallas, J.A.C.: Simulating memory effects with discrete dynamical systems. {Physica A} \textbf{198}, 339--339 (1993) (erratum)

\bibitem {ME1} Edelman, M.: On the fractional Eulerian numbers and equivalence of maps with long term power-law memory (integral
Volterra equations of the second kind) to Gr$\ddot{u}$nwald-Letnikov
fractional difference (differential) equations. {Chaos} \textbf{25},
073103 (2015)

\bibitem{ZT1}
Tarasov, V.E., Zaslavsky, G.M.:
Fractional equations of kicked systems and discrete maps.
{J. Phys. A} \textbf{41}, 435101 (2008)


\bibitem {MM7} Stanislavsky, A.A.:
Long-term memory contribution as applied to the motion 
of discrete dynamical system. 
{Chaos} \textbf{16}, 043105 (2006)

\bibitem{ME2} Edelman, M., Tarasov, V.E.:
Fractional standard map.
{Phys. Let. A} \textbf{374}, 279--285 (2009)

\bibitem{ME3} Edelman, M.:
Fractional standard map: Riemann-Liouville vs. Caputo.
{Commun. Nonlin. Sci. Numer. Simul.} \textbf{16}, 4573--4580 (2011)

\bibitem{ME4} Edelman, M., Taieb, L.A.:
New types of solutions of non-linear fractional differential
equations. In: Almeida, A., Castro, L, Speck F.-O. (eds.)
Advances in Harmonic Analysis and Operator Theory;
Series: Operator Theory: Advances and Applications
\textbf{229}, 139--155. Springer, Basel (2013)

\bibitem{ME5} Edelman, M.:
Universal fractional map and cascade of bifurcations type attractors. 
{Chaos} \textbf{23}, 033127 (2013)

\bibitem{ME6} Edelman, M.:
Fractional maps as maps with power-law memory.
In: Afraimovich, V., Luo, A.C.J.,Fu, X. (eds.)
{Nonlinear Dynamics and Complexity}, 79--120. Springer, New York (2014)

\bibitem{XML} Xiao, H., Ma, Y., Li, C.: Chaotic vibration in fractional maps. Journal of Vibration and Control \textbf{20}, 964--972 2014. DOI:10.1177/1077546312473769


\bibitem {GZ} Gray, H.L., Zhang, N.F.:
On a new definition of the fractional difference.
{Mathematics of Computation} \textbf{50}, 513--529 (1988)

\bibitem {MR} Miller, K.S., Ross, B.:
Fractional difference calculus.
In:  Srivastava H.M., Owa S. (eds.)
{Univalent Functions, Fractional Calculus, and Their Applications},    139--151. Ellis Howard, Chichester (1989)

\bibitem {Agr} Agarwal, R.P.: Difference equations and inequalities. Marcel Dekker, New York (2000)

\bibitem {GP} Goodrich, C., Peterson, A.: Discrete Fractional Calculus. Springer, New York (2015)

\bibitem{HBV2} Edelman, M:
Maps with power-law memory: direct introduction and Eulerian numbers,
fractional maps, and fractional difference maps.
In: Kochubei, A. and Luchko, Yu (eds.) {Handbookp
of Fractional Calculus With Applications}: Theory, vol. 2, pp. 47--64.
De Gruyter, Berlin (2019)

\bibitem{HBV4} Edelman, M.: Dynamics of nonlinear systems with power-law
memory. In: Tarasov, V.E. (ed.) Handbook of Fractional Calculus with Applications:  Applications in Physics, vol. 4, pp. 103--132. De Gruyter, Berlin (2019)


 

\bibitem {AE1} Atici, F.M., Eloe, P.W.: Initial value problems in discrete fractional calculus, Proc. Am. Math. Soc. \textbf{137}, 981--989 (2009)

\bibitem {Anast1} Anastassiou, G.A.: Discrete Fractional Calculus and Inequalities, http://arxiv.org/abs/0911.3370 (2009)

\bibitem {AE2} Atici, F. M., Eloe, P. W.: Discrete fractional calculus with the nabla operator. Electron. J. Qual.
Theory Differ. Equ., Spec. Ed. I. {2009}, No. 3, 1--12 (2009)

\bibitem {Anast2} Anastassiou, G.A.: Nabla discrete fractional calculus and nabla inequalities. Mathematical and Computer Modelling
\textbf{51}, 562--571 (2010).
DOI:10.1016/j.mcm.2009.11.006

\bibitem {BFT} Bastos, N. R. O., Ferreira, R. A. C., Torres, D. F. M.: Discrete-time fractional variational problems, Signal Process. \textbf{91}, 513--524 (2011)




\bibitem {CLZ} Chen, F., Luo, X., Zhou, Y.: Existence Results for Nonlinear Fractional Difference Equation. Adv.
Differ. Equ. \textbf{2011}, 713201 (2011)

\bibitem {WB1} Wu, G.-C., Baleanu, D., Zeng, S.-D.: Discrete chaos in fractional sine and standard maps. Phys. Lett. A
\textbf{378}, 484--487 (2014)

\bibitem {WB2} Wu, G.-C., Baleanu, D.: (2014), Discrete fractional logistic map and its chaos. Nonlin. Dyn. \textbf{75}, 283--287 (2014)

\bibitem {WB2C} Peng, Y., Sun, K., He, S., Wang, L.: Comments on Discrete fractional logistic map and its chaos. Nonlinear Dyn. 75, 283--287 (2014). Nonlin. Dyn. \textbf{97}, 897--901 (2019)



\bibitem {ME7} Edelman, M.:
Caputo standard $\alpha$-family of maps: Fractional difference vs. fractional.
{Chaos} \textbf{24}, 023137 (2014)

\bibitem{ME8} Edelman, M.:
Fractional maps and fractional attractors. Part ii: Fractional
difference $\alpha$-families of maps.
{Discontinuity, Nonlinearity, and Complexity} \textbf{4}, 391--402 (2015)

\bibitem {ME9} Edelman, M.:
On stability of fixed points and chaos in fractional systems.  
{Chaos} \textbf{28}, 023112 (2018)

\bibitem {ME10} Edelman, M.:
Universality in systems with power-law memory and fractional dynamics.
In: Edelman, M., Macau, E., Sanjuan, M.A.F. (eds.)
{Chaotic, Fractional, and Complex Dynamics: New Insights and Perspectives}. Ser.: Understanding Complex Systems, 147--171. Springer, eBook (2018)

\bibitem{ME12} Edelman, M.:
Cycles in asymptotically stable and chaotic fractional maps.
{Nonlinear Dynamics} \textbf{104}, 2829--2841 (2021)


\bibitem{ME13}
Edelman, M., Helman, A.B.: Asymptotic cycles in fractional maps of arbitrary positive orders. Fract. Calc. Appl. Anal. \textbf{25}, 181--206 (2022). DOI: 10.1007/s13540-021-00008-w

\bibitem{ME14} Edelman, M., Helman, A. B., Smidtaite, R.: Bifurcations and transition to chaos in generalized fractional maps of the orders     $0 < \alpha < 1$. Chaos \textbf{33}, 063123 (2023).
DOI: 10.1063/5.0151812


\bibitem{ME15} Edelman, M.: Stability of fixed points in generalized fractional maps of the orders $0 < \alpha < 1$. Nonlinear Dynamics, \textbf{111}, 10247--10254 (2023). DOI: 10.1007/s11071-023-08359-0


\bibitem{Elaydi} Elaydi, S.: An Introduction to Difference Equations. Undergraduate Texts in Mathematics. Springer, New York (2005)
 
\bibitem{ME16} Edelman, M.: Comments on A note on stability of fractional
logistic maps. Appl. Math. Lett. \textbf{129}, 107892 (2022)


\bibitem{Henon} H\'enon, M.: A two-dimensional mapping with a strange attractor. Communications in Mathematical Physics \textbf{50}, 69--77 (1976). DOI:10.1007/BF01608556 


\bibitem{Lozi} Lozi, R.: Un attracteur estrange (?) du type attracteur de H'enon. J. Physique (Paris) \textbf{39}, (Coll. C5), no. 8, 9--10 (1978)


\bibitem{FDSM} Edelman, M., Tarasov, V.E.: Fractional dissipative standard map. Chaos \textbf{20}, 023127 (2010). https://doi.org/10.1063/1.3443235

\bibitem{TarasovHenon} Tarasov, V.E.: Fractional Zaslavsky and H\'enon Discrete Maps. In: Luo, A.C.J., Afraimovich, V. (eds.)
{Long-range Interaction, Stochasticity and Fractional Dynamics}. Ser.: Nonlinear Physical Science, 1--26. Springer, Berlin, Heidelberg (2010). https://doi.org/10.1007/978-3-642-12343-6\_1 

\bibitem{KST} Kilbas, A.A., Srivastava, H.M., Trujillo, J.J.:
Theory and Application of Fractional Differential Equations.
Elsevier, Amsterdam (2006)

\bibitem{Podlubny} Podlubny I.: Fractional differential equations. Academic Press, San Diego (1999)

\bibitem{SKM} Samko, S.G., Kilbas, A.A., Marichev, O.I.: Fractional Integrals and Derivatives Theory and Applications. Gordon and Breach, New York (1993)

\bibitem{Hu} Hu, T.C.: Discrete chaos in fractional Henon map.
Applied Mathematics \textbf{5}, 2243--2248 (2014).
http://dx.doi.org/10.4236/am.2014.515218

\bibitem{LoziBif} Botella-Soler, V. et al.: Bifurcations in the Lozi map. Journal of Physics A: Mathematical and Theoretical \textbf{44}, 305101 (2011).
DOI:10.1088/1751-8113/44/30/305101


\bibitem{FrLoziBif} Khennaoui, A.-A. et al.:
On fractional--order discrete--time systems: Chaos, stabilization and synchronization. Chaos Solitons Fractals \textbf{119}, 150--162 (2019).
https://doi.org/10.1016/j.chaos.2018.12.019




\end{thebibliography}
\end{document}